\documentclass[prl,twocolumn,showpacs,preprintnumbers,amsmath,amssymb,superscriptaddress]{revtex4}
\usepackage{epsf}
\usepackage{graphicx}
\usepackage{bm}
\usepackage{amsmath}
\usepackage{color}
\usepackage{notes2bib}
\usepackage{epstopdf}
\usepackage{textcomp}

\bibnotesetup{
note-name = , use-sort-key = false }
\begin{document}

\title{Hall effect driven by non-collinear magnetic polarons in diluted magnetic semiconductors}

\author{K.~S.~Denisov}
\email{denisokonstantin@gmail.com} \affiliation{Ioffe Institute, 194021 St.Petersburg, Russia}
\affiliation{Lappeenranta University of Technology, FI-53851 Lappeenranta, Finland}
\author{N.~S.~Averkiev}
\affiliation{Ioffe Institute, 194021 St.Petersburg, Russia}

\begin{abstract}
In this letter we develop the theory of Hall effect driven by non-collinear magnetic textures (topological Hall effect - THE) in diluted magnetic semiconductors (DMS). 
We show that a carrier spin-orbit interaction induces a chiral magnetic ordering inside a bound magnetic polaron (BMP). 
The inner structure of non-collinear BMP is controlled by the type of spin-orbit coupling, allowing to create skyrmion- (Rashba) or antiskyrmion-like (Dresselhaus) configurations. 
The asymmetric scattering of itinerant carriers on polarons leads to the Hall signal which exists in weak external magnetic fields and low temperatures. 
We point out that DMS-based systems allow one to investigate experimentally the dependence of THE both on a carrier spin polarization and on a non-collinear magnetic texture shape. 
\end{abstract}


\date{\today}

\maketitle

The non-collinear ordering of magnetic moments is accompanied in experiment by an additional Hall signal (topological Hall effect - THE).
This phenomena has been evidenced in numerous magnetic systems with diverse shapes of non-collinear magnetization; the examples comprise spin glasses~\cite{SpinGlass1,SpinGlass2}, antiferromagnets~\cite{surgers2014large,ueland2012controllable}, Eu-based~\cite{EuO,ahadi2017evidence} and frustrated systems~\cite{Taguchi_Science,Machida_PRL}, topological insulators~\cite{THE_TI} and various ferromagnets with magnetic skyrmions~\cite{Muhl_MnSi_Science,MnSiAPhase,THE_Li,Chapman_PRB,Munzer_PRB,FeGe_THE}. 
It is now believed that the effect microscopically stems from a carrier exchange interaction with triads of non-collinear spins~\cite{Ye1999,Lyana-Geller1,Burkov,AronzonRozh,Tatara,prl_skyrmion,SciRep_skyrmion}. 
At the same moment, the theory predicts the different behavior of the emerging Hall response at strong and weak exchange couplings~\cite{SciRep_skyrmion}. 
In case of strong exchange interaction the transverse electric signal depends on a carrier spin polarization~\cite{BrunoDugaev,SciRep_skyrmion,Spin_Top_Hall}, while in the opposite regime of weak coupling THE appears even for totally unpolarized carriers~\cite{prl_skyrmion,SciRep_skyrmion,Tatara}.
The direct experiments focused on this issue have not been carried out so far, which is mainly due to the poor tuneability of carrier spin in those materials. 
To overcome this obstacle we propose to investigate the Hall effect driven by non-collinear magnetic textures in diluted magnetic semiconductors (DMS),
where the manipulation of carrier spin polarization is possible.

We describe a new mechanism for a non-collinear ordering of magnetic moments relevant for disordered paramagnetic systems, such as DMS. 
We demonstrate that a spin-orbit splitting of a carrier band states leads to a non-collinear internal pattern of a bound magnetic polaron (BMP) - a correlated state of a localized carrier and magnetic impurities lying inside a localization core. 
The non-collinear BMP favors asymmetric scattering of itinerant carriers and leads to an additional Hall response. 
The effect exists upon weak external magnetic fields up to the helium temperature and it carries an information about the type of a given spin-orbit interaction affecting the inner structure of BMP.

The key ingredient underlying the physics of non-collinear structures 
in disordered magnetic systems is an exchange interaction between carriers and magnetic impurities:
\begin{equation}\label{eq_Ex}
H_{ex} = - \alpha_{ex} \sum_n \left( {\boldsymbol{S}} \cdot {\boldsymbol{I}}_{n} \right) \hat{\rho}(\boldsymbol{r} - \boldsymbol{r}_n)
\end{equation}
where $\alpha_{ex}$ is an exchange coupling constant, $\hat{\rho}(\boldsymbol{r} - \boldsymbol{r}_n) = \delta\left(\boldsymbol{r} - \boldsymbol{r
}_n \right)$ is the density operator, 
${\boldsymbol{S}}$ is the operator of carrier spin (in zinc-blend semiconductors $S=1/2$ for $\Gamma_6$ conductivity band, $S=3/2$ for $\Gamma_8$ valence band), 
${\boldsymbol{I}}_n$ is the spin operator of magnetic impurity at point $\boldsymbol{r}_n$. 
On the one hand the exchange interaction (\ref{eq_Ex}) couples non-collinear magnetic pattern of $\boldsymbol{I}_n$ with a carrier orbital motion leading to Hall effect. On the other hand it allows a carrier to polarize magnetic moments along its spin, which results in a formation of a magnetic polaron. 

We provide the consideration for the zinc-blend A$_2$B$_6$ DMS quantum wells (QWs) doped by Mn. 
Let us describe the formation of non-collinear BMP. 
For this purpose we need to take into account the effect of a carrier on magnetic impurities. 
BMP is usually bonded to a localizing defect (donor or acceptor), so the coordinate part of a carrier wave function $\Psi_{\nu}(\boldsymbol{r})$ ($\nu = 1,2$ attributes to Kramers doublet) is controlled by the potential of an impurity center; the radius vector $\boldsymbol{r} = (\boldsymbol{\rho},z)$, where $\boldsymbol{\rho}$ is the coordinate in QW plane and $z$ is the coordinate along the QW growth axis. A carrier in state $\Psi_{\nu}$ generates an exchange field $\boldsymbol{B}_{ex}^{\nu}(\boldsymbol{r}_n)$ acting on magnetic impurity $\boldsymbol{I}_n$ at point $\boldsymbol{r}_n$. 
The energy of polaron $E_p^{\nu}$ in state $\nu$ upon the external magnetic field $\boldsymbol{B}_0$ is given by:
\begin{equation}
\label{eq_En_m}
E_{p}^{\nu} = - \sum_n g_{0} \mu_B {\boldsymbol{I}}_n \left( \boldsymbol{B}_{ex}^{\nu}(\boldsymbol{r}_n)   + \boldsymbol{B}_0 \right)
\end{equation}
where $g_0 =2$ is Mn gyromagnetic factor, $\mu_B$ is Bohr magneton and the sum goes over the impurities inside a localization radius. 
At low temperatures BMP falls into a collective state with magnetic moments ${\boldsymbol{I}}_n$ co-aligned with a direction of total magnetic field $\boldsymbol{B}_{ex}^{\nu}(\boldsymbol{r}_n) + \boldsymbol{B}_0$~\cite{Merkulov_Kavokin_2D,Berk},
which minimizes the energy $E_p^{\nu}$. 

The ordering of magnetic impurities $\boldsymbol{I}_n$ inside a BMP core in the collective regime is susceptible to the structure of local exchange field. To calculate $\boldsymbol{B}_{ex}^{\nu}(\boldsymbol{r})$ we average the exchange interaction (\ref{eq_Ex}) with a carrier wave function $\Psi_{\nu}(\boldsymbol{r})$ and compare the result with the energy in  (\ref{eq_En_m}). We get that the exchange field is given by:
\begin{equation}
\label{eq_Bex}
\begin{aligned}
&\boldsymbol{B}_{ex}^{\nu}(\boldsymbol{r}) = \frac{\alpha_{ex}}{g_{0} \mu_B} \Psi_{\nu}^{\dagger}(\boldsymbol{r}) \boldsymbol{S} \Psi_{\nu}(\boldsymbol{r}).
\end{aligned}
\end{equation}
In absence of carrier spin-orbit interaction the spinor part of $\Psi_{\nu}$ is a constant and the direction of exchange field $\boldsymbol{B}_{ex}^{\nu}$ is homogeneous in space. In this case magnetic moments inside a polaron core are all so-aligned~\cite{Merkulov_Kavokin_2D}. 
In presence of spin-orbit interaction a carrier spin is not conserved any longer~\cite{SmirnovGolub,kavokin2008spin}; the direction of vector 
$\boldsymbol{B}_{ex}^{\nu}(\boldsymbol{r})$ depends on coordinate~\cite{Kavokin-Sym,Kavokin-Asym,Berk} which leads to non-collinear magnetic ordering inside BMP.

We further consider the case when a spin-orbit interaction induces the rotation of spins within a QW plane:
\begin{equation}
\label{eq_I}
\boldsymbol{I}(\boldsymbol{r}) =  \left(I_{\parallel}(\rho,z)\cos{\Phi(\theta)} , I_{\parallel}(\rho,z)\sin{\Phi(\theta)}, I_z(\rho,z)\right)
\end{equation}
where $\boldsymbol{\rho} = (\rho,\theta)$ and $\rho$ is the QW in-plane radius-vector magnitude, $\theta$ is the polar angle, 
$I_{\parallel},I_z$ are in- and out of-plane spin components respectively, the function $\Phi(\theta) = (\varkappa \theta + \gamma)$ describes the in-plane spin rotation. The vorticity $\varkappa$, which takes integer values, and the initial phase of rotation $\gamma$ depend on a particular type of spin-orbit coupling. 
The dependence of BMP spins $\boldsymbol{I}$ on $z$ might arise from an impurity confinement (QW thickness $d_{QW}$ is larger than a Bohr radius $a_B$, e.g. the acceptor polaron from $\Gamma_8$ valence band~\cite{Merkulov_Polaron,Berk}), or simply reflect the structure of a carrier envelop function along $z$ (narrow QWs with $d_{QW}<a_B$).

To describe the appearance of Hall effect in QW due to non-collinear BMPs let us consider the motion of an itinerant carrier (either electron or hole) with $1/2$ pseudo-spin $\boldsymbol{j}$ in the frozen spin field $\boldsymbol{I}(\boldsymbol{r})$ of an individual polaron (meanfield approximation). 
After integrating $\boldsymbol{I}(\boldsymbol{r})$ in (\ref{eq_Ex}) with itinerant carrier subband envelop function along $z$ and taking into account that ${\boldsymbol{S}}_{\alpha} = {g}_{\alpha \beta} \boldsymbol{j}_{\beta}$
(tensor ${g}_{\alpha \beta}$ couples the matrix elements of $\boldsymbol{S}$ and $\boldsymbol{j}$ operators)
we get the $2\times 2$ matrix scattering potential within carrier subband states:
\begin{equation}
\label{eq_Vsk}
V_{sc}(\boldsymbol{\rho}) = -\frac{1}{2}\alpha_{ex} n_m
\begin{pmatrix}
L_{z}(\rho) & e^{- i \varkappa \theta - i \gamma} L_{\parallel}(\rho) 
\\
e^{ i \varkappa \theta + i \gamma} L_{\parallel}(\rho) & -L_{z}(\rho)
\end{pmatrix},
\end{equation}
where $n_m = x_{\rm Mn} N_0$ is a density of magnetic impurities ($N_0$ is a number of cations per unit cell, $x_{\rm Mn}$ is a fraction of Mn), 
$L_{z} = \langle I_{z}(\rho) \rangle g_{z}$, $L_{\parallel} = \langle I_{\parallel}(\rho) \rangle g_{\parallel}$, and $g_{\parallel}, g_z$ are in- and out-of-plane components of $g_{\alpha \beta}$. 
The in-plane spin rotation of $\boldsymbol{I}(\boldsymbol{r})$ enters in the off-diagonal components of the scattering potential via the angular factors $e^{\pm i \varkappa \theta}$ and gives rise to a carrier scattering asymmetry~\cite{SciRep_skyrmion}. 
There must be $g_{\parallel} \neq 0$ to preserve this asymmetry so the effect might be weakened for heavy hole like QW subbands.

The Hall effect due to a carrier asymmetric scattering on BMPs has a number of features \cite{SciRep_skyrmion}. Let us focus on those that arise from the polaron-based mechanism of spin non-collinearity. 
The non-collinear ordering inside BMP 
is destroyed both by strong external magnetic field (it tilts all spins along its direction) and by high temperature due to thermal fluctuations.
Thus the observation of THE requires low temperatures and weak external magnetic fields. 
Besides, the sign of Hall signal depends on a polaron vorticity $\varkappa$~\cite{SciRep_skyrmion}, which is determined by the type of carrier spin-orbit splitting and takes different values. Let us consider these issues in details.

We consider BMP in an ultra-narrow QW ($d_{QW}<a_B$). 
We focus on a purely 2D model with the exchange field $\boldsymbol{B}_{ex}(\boldsymbol{\rho})$ depending on QW plane radius vector $\boldsymbol{\rho}$. 
Let a 2D polaron acquires a non-collinearity due to 
linear in $\boldsymbol{k}$ Rashba or Dresselhaus spin-orbit subband splittings of a carrier with $1/2$ pseudo-spin~\cite{DenisovPolaron}. 
In the coordinate system $x||[100], y||[010]$, $z || [001]$ (QW is grown along $z$-axis) the $\boldsymbol{k}$-dependent spin-orbit part of carrier Hamiltonian is given by:
\begin{equation}
\label{eq_SOI}
H_{SO}(\boldsymbol{k}) = \beta_{SO} k \begin{pmatrix}
0 & e^{-i\chi\theta_{{k}} - i\gamma_k} \\
e^{i\chi\theta_{{k}}+ i\gamma_k} & 0
\end{pmatrix}
\end{equation}
where $\beta_{SO}$ is a spin-orbit coupling, $\boldsymbol{k} = (k,\theta_k)$ is a planar wave-vector ($k$ - magnitude and $\theta_{k}$ - polar angle); $\chi = 1$ for Rashba and $\chi= -1$ for Dresselhaus splitting, and $\gamma_k = -(1+\chi) \pi/4 $. 
We point out the similarity between $H_{SO}(\boldsymbol{k})$ and $V_{sc}(\boldsymbol{\rho})$. 
The spin-orbit Hamiltonian $H_{SO}$ describes spin rotation in ${\boldsymbol{k}}$-space with the direction determined by $\chi$, while $V_{sc}$ describes spin rotation in $\boldsymbol{\rho}$-space with direction given by $\varkappa$. Naturally, the type of an initial spin-orbit interaction $\chi = \pm 1$ is inherited by the type of BMP $\varkappa = \chi$.

Let a particle (either electron or hole) from the lowest QW subband affected by $H_{SO}$  is bounded by an impurity potential $V_0(\rho)$.  
The wave function $\Psi_{\nu}(\boldsymbol{r}) = \psi_{\nu}(\boldsymbol{\rho})/\sqrt{d_{QW}}$ 
contains a planar component $\psi_{\nu}(\boldsymbol{\rho})$ which  satisfies the Schrodinger's equation: 
\begin{equation}
\label{eq_Hw}
\left( -\frac{\hbar^2}{2m}\nabla^2 + H_{SO}+ V_0(\rho) - E_0\right) \psi_{\nu}(\boldsymbol{\rho}) = 0 
\end{equation}
where the first term attributes to a parabolic free-motion dispersion of 2D subband with in-plane effective mass $m$, 
and $E_0$ is an energy of bound state counted from a subband edge at $\beta_{SO} =0$. 
Although there is a series of localized states in 2D system with spin-orbit splitting~\cite{ChaplikPRL}, we consider only the lowest Kramers pair taking its energy $E_0$ as a parameter ($E_0 < -m \beta_{SO}^2/2\hbar^2$). 
We also neglect the effect of the weak external magnetic field. 
For a short-range potential $V_0(\rho) \sim \delta(\boldsymbol{\rho})$ the solutions $\psi_{\nu}(\boldsymbol{\rho})$ of (\ref{eq_Hw}) are given by~\footnote{The details of calculations are given in the Supplementary Materials, Appendix A}:
\begin{equation}
\label{eq_Wave}
\psi_1(\boldsymbol{\rho}) = 
\begin{pmatrix}
a(\rho)
\\
{e^{ i \varkappa \theta + i \gamma'} } b(\rho)
\end{pmatrix},
\hspace{0.4cm}
\psi_2(\boldsymbol{\rho}) = - i \sigma_2 \psi_1^{\ast}(\boldsymbol{\rho})
\end{equation}
where $\theta$ is measured from $x||[100]$, 
$\sigma_2$ is the second Pauli matrix, $\gamma' = (1-\varkappa) \pi/4$, the functions $a(\rho) = c_0 {\rm Re}\left[q K_0(q\rho)\right]$, $b(\rho) = c_0 {\rm Im}\left[q K_1(q\rho) \right]$, 
$q = q_0 \left( \sqrt{1 - \delta_{SO}^{2}} + i \delta_{SO} \right)$, 
$q_0 = \sqrt{2 m |E_0|/ \hbar^2}$, 
$q_{SO} = m \beta_{SO}/\hbar^2$, 
$\delta_{SO} = q_{SO}/q_0$,
$K_0(z), K_1(z)$ are modified Bessel functions of zero-th and first kind respectively, and the normalization constant $c_0 = \sqrt{\pi } \left(1 - \ln{q/q^{\ast}} (q-q^{\ast})/2(q + q^{\ast}) \right)^{-1/2}$. 
The vorticity $\varkappa=\pm 1$ in (\ref{eq_Wave}) attributes either to Rashba or Dresselhaus spin-orbit interaction.

Using the wave functions $\psi_{\nu}(\boldsymbol{\rho})$ from (\ref{eq_Wave}) we obtain a general form of the exchange field:
\begin{equation}
\label{eq_j}
\begin{aligned}
& \boldsymbol{B}_{ex}^{x}(\boldsymbol{\rho}) = \zeta g_{\parallel} a(\rho) b(\rho) \cos{\left( \varkappa \theta + \gamma\right)}
\\
&\boldsymbol{B}_{ex}^{y}(\boldsymbol{\rho}) = \zeta g_{\parallel} a(\rho) b(\rho) \sin{\left( \varkappa \theta + \gamma \right)}\\
& \boldsymbol{B}_{ex}^{z}(\rho) = \zeta g_z \frac{(-1)^{\nu}}{2} \left( b^2(\rho) - a^2(\rho) \right)
\end{aligned}
\end{equation}
where $\zeta = (\alpha_{ex}/{g_0 \mu_B d_{QW}} ) $,  $\gamma$ is an initial phase. 
As soon as $g_{\parallel} \neq 0$ the exchange field 
gets the in-plane rotation and induces a chiral ordering of $\boldsymbol{I}_n$ inside a polaron core. The latter favors an asymmetric carrier scattering via the potential $V_{sc}$ from (\ref{eq_Vsk}). The sign of an emerging scattering asymmetry depends on $\varkappa = \pm 1$, thus Rashba and Dresselhaus magnetic polarons give rise to an observable Hall effect of the opposite sign. 


\begin{table}
	\begin{tabular}[t]{|c|c|c|}
		\hline
		$\gamma$  & $\varkappa = +1 $ & $\varkappa = -1 $\\
		\hline
		$\eta = 1 $ & $ 0 $ & $ {\pi}/{2} $  \\
		\hline
		$\eta = -1$ & $ {\pi}$   & $ \displaystyle -{\pi}/{2} $  \\
		\hline
	\end{tabular}		
	\caption{The polaron in-plane spin rotation phase $\gamma$ for different orientation $\eta=\pm 1$ and vorticity $\varkappa$.}
	\label{tab_1}
\end{table}

Let the external magnetic field $\boldsymbol{B}_0$ is directed along the QW growth axis. 
The Kramers states $\nu$ describe two polaron configuration with opposite orientation of spins $\eta = \pm 1$ in the very center with respect to $\boldsymbol{B}_0$. 
The parallel configuration $\eta =1$ has the shape of a non-collinear ring; the anti-parallel $\eta=-1$ describes a topologically charged state called magnetic skyrmion for $\varkappa = 1$ or antiskyrmion $\varkappa = -1$. The presence of $\boldsymbol{B}_0$  lifts the Kramers degeneracy and $\eta=\pm 1$ polaron configurations have different energies (\ref{eq_En_m}); the one with $\eta=1$ tends to be the ground state. 
The initial phase of in-plane rotation $\gamma$ depends both on $\eta$ and $\varkappa$, which is given in Table (\ref{tab_1}). 

\begin{figure}
	\centering	
	\includegraphics[width=0.5\textwidth]{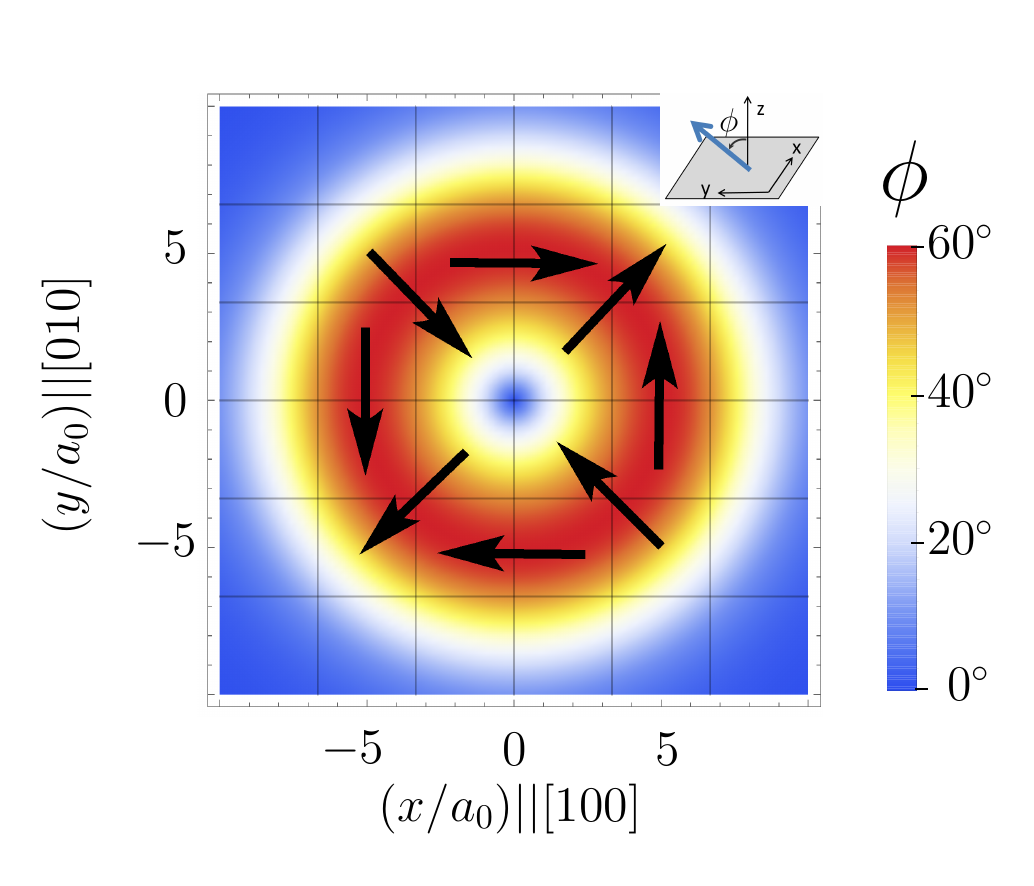}
	\caption{
		Bound magnetic polaron formed by a light hole with Dresselhaus spin-orbit band splitting. 
		The color shows the magnitude of $\boldsymbol{I}$ inclination angle $\phi$ at different distance from an acceptor center. 
		$E_0 = - 20$ meV, $m = 0.4 m_0$, $\beta_{SO} = 250$ meV$\mathring{A}$ ($\delta_{SO} = 0.3$), $B_0 =50$ mT, $\varkappa = -1$, $\gamma = \pi/2$, $a_0=0.64$ nm, the other parameters are taken for CdTe}	
	\label{fig1}
\end{figure}

At low temperatures $T \rightarrow 0$ a BMP falls into fully saturated collective regime when the magnetic moments inside BMP core are so-aligned with a direction of total magnetic field:
\begin{equation}
\label{eq_Ir}
\boldsymbol{I}(\boldsymbol{\rho}) = I \frac{\boldsymbol{B}_{ex}(\boldsymbol{\rho}) + \boldsymbol{B}_0}{|\boldsymbol{B}_{ex}(\boldsymbol{\rho}) + \boldsymbol{B}_0|}
\end{equation}
where $I$ is a magnitude of spin
(we assume that there is a big number of magnetic impurities inside BMP $x_{\rm Mn} N_0 \pi a_B^2 d_{QW} \gg 1$).
In Fig.\ref{fig1} we show the distribution of spins $\boldsymbol{I}(\boldsymbol{\rho})$ inside a BMP with $\eta=1$
created by a light hole (effective mass $m = 0.4 m_0$) bound to an acceptor ($E_0 =- 20$ meV)~\footnote{
	We consider the case of tensed QW with heavy hole subband 
pushed deep down. }
in a narrow QW ($d_{QW} = 2$ nm) with the Dresselhaus spin-orbit interaction~\cite{Durnev_Glazov} ($\beta_{SO} = 250$ meV$\mathring{A}$, $\delta_{SO} = 0.3$).  
The exchange coupling constant in Fig.\ref{fig1} is taken for CdMnTe $\alpha_{ex}  = \alpha_0/3$, where $\alpha_0 N_0 = -880$ meV, 
$I=5/2$ for Mn, the magnetic field $B_0 = 50$ mT.  
The spatial coordinates are given in units of lattice constant $a_0 = 0.64$ nm. 
The components of light hole g-factor are $g_{\parallel} = 2$, $g_z =1$. 
The polaron in Fig.\ref{fig1} has a shape of non-collinear ring with a pronounced magnetization tilt (the maximum value $\phi_{max} \approx 60^\circ$) at $2-4$ nm from center. 
The in-plane spin rotation is determined by Dresselhaus spin-orbit interaction ($\varkappa = -1$, $\gamma = \pi/2$). 

\begin{figure}
	\centering	
	\includegraphics[width=0.5\textwidth]{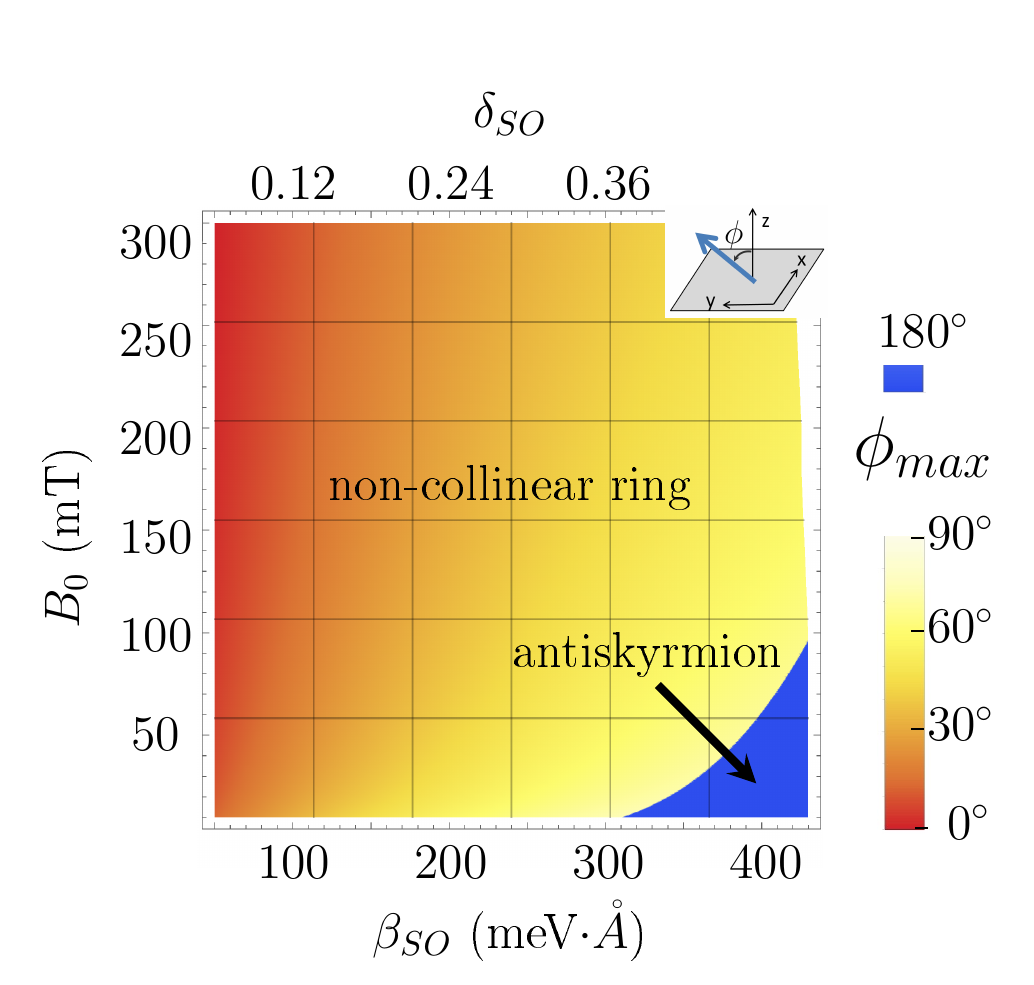}
	\caption{The map of maximum tilt angles $\phi_{max}$ inside a BMP at ground state. The region of $\phi_{max} = \pi$ ($\delta_{SO}>0.35$ and weak $B_0$) corresponds to a skyrmion ($\varkappa = 1$) or antiskyrmion ($\varkappa = -1$)  configuration at $T\to 0$. The parameters are the same as in Fig.\ref{fig1}.}
	\label{fig2}
\end{figure}

A non-collinear polaron structure disappears with increase of external magnetic field $B_0$ or temperature $T$. 
The suppression occurs when $B_0$, $T$ become comparable with the magnitude of exchange field in the region where the spin-orbit interaction induces in-plane twist of $\boldsymbol{I}$. 
For the considered example of Rashba and Dresselhaus 2D polarons the spatial position of this region is controlled by the parameter $\delta_{SO} =q_{SO}/q_0$ ($\boldsymbol{B}_{ex}^{\parallel} \sim a(\rho) b(\rho)$, where $b(\rho) \sim {\rm Im}\left[q K_1(q\rho) \right]$ is nonzero due to $\delta_{SO}\neq 0$).  
The exchange field of a localized carrier decreases away from a polaron center, thus 
to shift the non-collinear ring closer to polaron center the larger spin-orbit coupling (in 2D case larger $\delta_{SO}$) is needed. 
In Fig.\ref{fig2} we show the map of maximum tilt angles $\phi_{\rm max}$ inside a BMP being at ground state for different $B_0$ and $\delta_{SO}$ at $T\to 0$ (the other parameters are the same as in Fig.\ref{fig1}). 
For small $\delta_{SO}$ a non-collinear region appears only outside the Bohr radius and it is destroyed by extremely weak $B_0$. 
At larger spin-orbit splitting ($\beta_{SO} = 250$ meV$\mathring{A}$, $\delta_{SO} =0.3$) 
the exchange field rotates already within the Bohr radius (see Fig.\ref{fig1}) with a spin tilt $\phi_{max} \approx 60^\circ$ at $B_0 = 50$ mT and $\phi_{max} \approx 15^\circ$ at $B_{0} = 0.3$ T. 
Alignment of spins $\boldsymbol{I}$ along the $\boldsymbol{B}_0$ diminishes the Hall effect.

Another property of 2D non-collinear BMP model is that at large $\delta_{SO}$ and $T\to 0$ the polaron ground state undergoes a transition to a skyrmion (antiskyrmion at $\varkappa = -1$) configuration with $\eta = -1$ (see the region of $\phi_{max} = -\pi$ in Fig.~\ref{fig2}).
The maximum gain in polaron energy $E_p^{\nu}$ (ground state) 
occurs when 
the QW perpendicular exchange field component $B_{ex}^z$ and $\boldsymbol{B}_0$ are so-aligned
for the majority of spins inside BMP, 
which is ordinarily achieved for $\eta=1$. 
Nevertheless, at large $\delta_{SO}$ the sign of $B_{ex}^z$ changes at some distance from polaron center.  
In this case it is more favorable to orient the periphery spins along the external field, while few spins in the center will be directed oppositely creating a skyrmion with $\eta=-1$~\footnote{The details are given in Supplementary Materials, appendix B.}.

The rotating BMP spin texture from (\ref{eq_I}) also appears in other DMS systems, e.g. in wide QWs with $\Gamma_8$-like acceptor polaron~\cite{Berk,Merkulov_Polaron}, or in structures with inverse spectrum (e.g. HgTe/CdTe QWs) ~\cite{Burmistrov}. 
This makes it possible to investigate the topological Hall response from a wide range of non-collinear magnetic configurations on the unified platform: the desirable magnetic texture can be achieved by the design of a heterostructure. 

The BMP-based THE in an arbitrary DMS system is distinguished from other transverse contributions (ordinary and anomalous Hall effects) by the existence only within the region where non-collinearity is preserved ($B_{0} \le (0.5-1)$T, $T \le (3-4) $K). 
Moreover, DMS-based heterostructures give control over the carrier spin polarization $P_s$ in this region. 
Changing the ratio $\mu_B g_0 B_0 I/kT$ we control the magnitude of background magnetization and regulate $P_s$ via the giant Zeeman effect. 
This allows to investigate experimentally the dependence of THE on carrier spin polarization. 
This dependence is a keystone of the phenomenon: the theory states the existence of different regimes with respect to $P_s$ \cite{SciRep_skyrmion}.

A carrier asymmetric scattering on BMP dramatically depends on the magnitude of adiabatic parameter $\lambda_a = \omega_{ex} \tau$~\cite{SciRep_skyrmion}, where $\hbar \omega_{ex}$ is an exchange interaction strength and $\tau$ is a characteristic time of a carrier fly through a texture core. 
At the adiabatic regime $\lambda_a \gg 1$ spin-up and spin-down carriers are scattered in the opposite transverse directions leading to spin Hall effect. 
In this regime the electrical transverse response is proportional to a carrier spin polarization $P_s$. 
In the opposite regime $\lambda_a \le 1$ a carrier is scattered disregard to the initial spin state to the unique transverse direction controlled by polaron orientation $\eta =\pm 1$ \cite{prl_skyrmion}. 
In this case the Hall signal can be observed even for totally unpolarized carriers $P_s = 0$ provided that there is nonequivalent number of $\eta = \pm 1$ polarons in the sample 
(upon increasing of ${B}_0$ the number of BMPs with $\eta=1$ prevails as they have lower energy). 
The crossover regime $\lambda_a \sim 1$ is accompanied by oscillations of Hall response upon varying a polaron size or Fermi energy. 
Carrying out the measurements of electrical transverse signal at different $\mu_B g_0 B_0I/kT$ gives us an experimental access to the physics of topological Hall effect in different regimes.

Let us comment the possible design of DMS-based heterostructure. We propose A$_2$B$_6$ based QW doped both by manganese and 
neutral acceptors $A^0$ with latter forming non-collinear BMPs. 
A conductive channel created by barrier doping might be located either within the same QW or behind a spacer, i.e. on the basis of double QW structure. 
We control $\lambda_a$ by means of varying the concentration of itinerant carriers (either electrons $n_e$ or holes $n_h$) and the fraction $x_{\rm Mn}$ of Mn.
For a single CdTe-based ultra-narrow QW ($d_{QW} = 2$ nm) structure with BMP of $6$ nm size the adiabatic regime is typical for 2DHG ($\lambda_a = 16$ for $x_{\rm Mn} = 0.1$, the light hole Fermi energy $E_F = 2.3$ meV, $n_h = 5 \times 10^{11}$ cm$^{-2}$, the effective mass $m_h = 0.4 m_0$), while weak coupling regime is achieved at small $x_{\rm Mn}$ for dense 2DEG ($\lambda_a = 0.7$ for $x_{\rm Mn}=0.035$, electron Fermi energy $E_F = 18.5$ meV, $n_e = 10^{11}$ cm$^{-2}$, $m_e = 0.1 m_0$). 
In double QW structures there is a decrease of an effective exchange coupling which pushes $\lambda_a$ towards crossover and weak coupling regime for both 2DHG and 2DEG.

Summarizing, we have proposed that in diluted magnetic semicondutors an additional Hall response is induced due to a carrier asymmetric scattering on non-collinear bound magnetic polarons. 
We call for the experiments with DMS-based QWs, which allow one to create diverse configurations of non-collinear BMPs and to study the dependence of emerging Hall effect on a carrier spin polarization.

We thank I.~V.~Rozhansky, Y.~G.~Kusraev, B.~G.~Namozov,  E.~Lahderanta for helpful comments. 
The work has been carried out under the financial support of
Grants from the Russian Science Foundation (analytical theory - project no.17-12-01182; physical analysis - project no.17-12-01265). 
K.S.D. and N.S.A. thank the Foundation for the Advancement of Theoretical Physics and Mathematics "BASIS", and the program of RAS \textnumero 9 THz optoelectronics and spintronics.


\bibliography{Skyrmion}

\end{document}